\documentclass{PoS}
\PoS{PoS(HEP2005)418}
\title{Short study of the $\eta$-$\eta^\prime$ system in the two mixing angle scheme}
\ShortTitle{Short study of the $\eta$-$\eta^\prime$ system in the two mixing angle scheme}
\author{\speaker{Rafel Escribano}
              \thanks{UAB--FT--592 report.
                            Work partly supported by the Ramon y Cajal program,
                            the Ministerio de Ciencia y Tecnolog\'{\i}a and FEDER, FPA2002-00748EU,
                            and the EU, HPRN-CT-2002-00311, EURIDICE network.}\\
              Grup de F\'{\i}sica Te\`orica and IFAE, Universitat Aut\`onoma de Barcelona,\\
              E-08193 Bellaterra (Barcelona), Spain\\
              E-mail: \email{Rafel.Escribano@ifae.es}}
\abstract{
A two mixing angle description of the pseudoscalar decay constants associated to the 
$\eta$-$\eta^\prime$ system is used to parametrize the theoretical amplitudes of the
radiative decays $(\eta,\eta^\prime)\to\gamma\gamma$ and the coupling constants
$g_{V(\eta,\eta^\prime)\gamma}$ with $V=\rho,\omega,\phi$.
The parametrization is performed in both the ``octet-singlet'' basis and the ``quark-flavour'' basis.
An excellent agreement with the most recent experimental data is achieved.
Our analysis reveals that at the present experimental accuracy the two mixing angles
differ significantly in the former basis but not in the latter,
in accordance with the expectations of large $N_c$ Chiral Perturbation Theory
where the difference between the two mixing angles are due to a $SU(3)_f$-breaking effect
and a violation of the OZI rule respectively. 
}
\FullConference{International Europhysics Conference on High Energy Physics\\
                              July 21st - 27th 2005\\
                              Lisboa, Portugal}
\begin{document}
\section{Theoretical input}
The decay constants of the $\eta$-$\eta^\prime$ system in the octet-singlet basis
$f_P^a\ (a=8,0; P=\eta,\eta^\prime)$ are defined as $\langle 0|A_\mu^a|P(p)\rangle=i f_P^a p_\mu$,
where $A_\mu^{8,0}$ are the octet and singlet axial-vector currents\footnote{
The axial-vector currents are defined as
$A_\mu^a=\bar q\gamma_{\mu}\gamma_{5}\frac{\lambda^a}{\sqrt{2}}q$
with the normalization convention $f_{\pi}=\sqrt{2}F_{\pi}=130.7$ MeV.}.
The divergence of the matrix elements are written as
$\langle 0|\partial^\mu A_\mu^a|P\rangle=f_P^a m_P^2$,
where $m_P$ is the mass of the pseudoscalar meson.
Each of the two mesons $P=\eta, \eta^\prime$ has both octet and singlet components, $a=8, 0$.
Consequently, the matrix elements define \emph{four independent} decay constants.
Following the convention of Ref.~\cite{Leutwyler:1997yr}
the decay constants are parametrized in terms of two basic decay constants
$f_{8}, f_{0}$ and two angles $\theta_{8}, \theta_{0}$:
\begin{equation}
\label{defdecaybasisos}
\left(
\begin{array}{cc}
f^8_\eta & f^0_\eta \\[1ex]
f^8_{\eta^\prime} & f^0_{\eta^\prime}
\end{array}
\right)
=
\left(
\begin{array}{cc}
f_8 \cos\theta_8 & -f_0 \sin\theta_0 \\[1ex]
f_8 \sin\theta_8 &  f_0 \cos\theta_0
\end{array}
\right)\ .
\end{equation}
Analogously to the $\pi^0\rightarrow\gamma\gamma$ case,
one assumes that the interpolating fields $\eta$ and $\eta^\prime$ can be
related with the axial-vector currents.
This leads to
\begin{equation}
\label{theowidths}
\Gamma(\eta,\eta^\prime\rightarrow\gamma\gamma)=
\frac{\alpha^2 m_\eta^3}{96\pi^3}\left(
\frac{c\theta_0/f_8-2\sqrt{2}s\theta_8/f_0}
{c\theta_0 c\theta_8+s\theta_8 s\theta_0}\right)^2, 
\frac{\alpha^2 m_{\eta^\prime}^3}{96\pi^3}\left(
\frac{s\theta_0/f_8+2\sqrt{2}c\theta_8/f_0}
{c\theta_0 c\theta_8+s\theta_8 s\theta_0}\right)^2\ .
\end{equation}

The electromagnetic form factors of the radiative decays of lowest-lying vector mesons,
$V\rightarrow(\eta,\eta^\prime)\gamma$, and of the radiative decays
$\eta^\prime\rightarrow V\gamma$, with $V=\rho, \omega, \phi$,
are described in terms of their relation with the $AVV$ triangle anomaly, $A$ 
and $V$ being an axial-vector and a vector current respectively.
The approach both includes $SU(3)_f$ breaking effects and fixes the vertex couplings 
$g_{VP\gamma}$ as explained below.
One considers the correlation function
$i\int d^4x e^{iq_1 x}\langle P(q_1+q_2)|TJ_\mu^{\rm EM}(x)J_\nu^V(0)|0\rangle=
   \epsilon_{\mu\nu\alpha\beta}q_1^\alpha q_2^\beta F_{VP\gamma}(q_1^2,q_2^2)$,
where the currents are defined as
$J_\mu^{\rm EM}=\frac{2}{3}\bar u\gamma_\mu u-\frac{1}{3}\bar d\gamma_\mu d-
                                 \frac{1}{3}\bar s\gamma_\mu s$, 
$J_\mu^{\rho,\omega}=\frac{1}{\sqrt{2}}(\bar u\gamma_\mu u\mp\bar d\gamma_\mu d)$ and
$J_\mu^\phi=-\bar s\gamma_\mu s$.
The form factors values $F_{VP\gamma}(0,0)$ are fixed by the $AVV$ triangle 
anomaly (one $V$ being an electromagnetic current), and are written in terms 
of the pseudoscalar decay constants and the $\phi$-$\omega$ mixing angle 
$\theta_V$ (see Ref.~\cite{Escribano:2005qq} for the complete expressions).
Using their analytic properties, we can express these form factors by dispersion relations
in the momentum of the vector current, which are then saturated with the lowest-lying resonances
$F_{VP\gamma}(0,0)=\frac{f_V}{m_V}g_{VP\gamma}+\cdots\,$,
where the dots stand for higher resonances and multiparticle contributions to the correlation function.
In the following we assume vector meson dominance (VMD) and thus neglect these contributions
(see Ref.~\cite{Ball:1995zv} for further details).
The $f_V$ are the leptonic decay constants of the vector mesons defined by
$\langle0|J_\mu^V|V(p,\lambda)\rangle=m_V f_V \varepsilon_\mu^{(\lambda)}(p)$,
where $m_V$ and $\lambda$ are the mass and the helicity state of the vector meson.
The $f_V$ can be determined from the experimental decay rates 
$\Gamma(V\rightarrow e^+e^-)=\frac{4\pi}{3}\alpha^2\frac{f_V^2}{m_V}c_V^2$, 
with $c_V=(\frac{1}{\sqrt{2}},\frac{s\theta_V}{\sqrt{6}},\frac{c\theta_V}{\sqrt{6}})$
for $V=\rho, \omega, \phi$.
Finally, we introduce the vertex couplings $g_{VP\gamma}$,
which are just the on-shell $V$-$P$ electromagnetic form factors:
$\langle P(p_P)|J_\mu^{\rm EM}|V(p_V,\lambda)\rangle|_{(p_V-p_P)^2=0}=
-g_{VP\gamma}\epsilon_{\mu\nu\alpha\beta}p_P^\nu p_V^\alpha\varepsilon_V^\beta(\lambda)$.
The decay widths are
$\Gamma(P\rightarrow V\gamma)=
\frac{\alpha}{8}g_{VP\gamma}^2\left(\frac{m_P^2-m_V^2}{m_P}\right)^3$ and
$\Gamma(V\rightarrow P\gamma)=
\frac{\alpha}{24}g_{VP\gamma}^2\left(\frac{m_V^2-m_P^2}{m_V}\right)^3$.

In the quark-flavour basis the decay constants are parametrized in terms of
$f_{q}, f_{s}$ and $\phi_{q}, \phi_{s}$:
\begin{equation}
\label{defdecaybasisqf}
\left(
\begin{array}{cc}
f^q_\eta & f^s_\eta \\[1ex]
f^q_{\eta^\prime} & f^s_{\eta^\prime}
\end{array}
\right)
=
\left(
\begin{array}{cc}
f_q \cos\phi_{q} & -f_s \sin\phi_{s}\\[1ex]
f_q \sin\phi_{q} &  f_s \cos\phi_{s}
\end{array}
\right)\ ,
\end{equation}
and the non-strange and strange axial-vector currents are defined as
$A_\mu^q=\frac{1}{\sqrt{3}}(A_\mu^8+\sqrt{2}A_\mu^0)$ and
$A_\mu^s=\frac{1}{\sqrt{3}}(A_\mu^0-\sqrt{2}A_\mu^8)$.
The parametrization of the decay widths $\Gamma(\eta,\eta^\prime\to\gamma\gamma)$
in the quark-flavour basis and the vertex couplings $g_{VP\gamma}$  in both basis are found in
Ref.~\cite{Escribano:2005qq}.

\section{Results}
In order to test our theoretical predictions with the most recent experimental information accounting for 
$(\eta,\eta^\prime\to\gamma\gamma)$ and the radiative vector decays
$V\to P\gamma$ and $P\to V\gamma$ \cite{Eidelman:2004wy}
we must first know the values of the mixing parameters (decay constants and mixing angles)
preferred by the data.
Therefore, we have performed various fits to this set of experimental data assuming, or not, 
the two mixing angle scheme of the $\eta$-$\eta^\prime$ system.
The results for the octet-singlet basis are presented in the upper part of Table \ref{table1}.
The theoretical constraint $f_8=1.28 f_\pi$ is relaxed in order to test the dependence of the result
on the value of this parameter.
The experimental constrain $\theta_V=(38.7\pm 0.2)^\circ$ is kept fixed in all the fits.
\begin{table}
\centering
{\footnotesize
\begin{tabular}{ccc|ccc}
\hline\hline
&&&&&\\
Assumptions & Results & $\chi^2/\textrm{d.o.f.}$ &
Assumptions & Results & $\chi^2/\textrm{d.o.f.}$\\[1ex]\hline
&&&&&\\
$\theta_8$ and $\theta_0$ free  & $\theta_8=(-22.2\pm 1.4)^\circ$   & 40.5/5&
$\theta_8=\theta_0\equiv\theta$ & $\theta=(-18.9\pm 1.2)^\circ$       & 73.7/6\\[1ex]
$f_8=1.28 f_\pi$            & $\theta_0=(-5.5\pm 2.3)^\circ$  &       &
$f_8=1.28 f_\pi$            & $f_0=(1.11\pm 0.03) f_\pi$        &   \\[1ex]
                                         & $f_0=(1.25\pm 0.04) f_\pi$        &       &
                                         &                   &   \\[1ex]\hline
&&&&&\\
$\theta_8$ and $\theta_0$ free  & $\theta_8=(-23.8\pm 1.4)^\circ$   & 17.9/4&
$\theta_8=\theta_0\equiv\theta$ & $\theta=(-18.2\pm 1.2)^\circ$       & 66.9/5\\[1ex]
$f_8$ free              & $\theta_0=(-1.1\pm 2.3)^\circ$  &       &
$f_8$ free              & $f_8=(1.39\pm 0.04) f_\pi$        &   \\[1ex]
                                & $f_8=(1.51\pm 0.05) f_\pi$        &       &
                                & $f_0=(1.13\pm 0.03) f_\pi$        &   \\[1ex]
                                & $f_0=(1.32\pm 0.05) f_\pi$        &       &
                                &                   &   \\[1ex]
\hline\hline
&&&&&\\
$\phi_q$ and $\phi_s$ free  & $\phi_q=(42.7\pm 2.0)^\circ$      & 32.6/5&
$\phi_q=\phi_s\equiv\phi$   & $\phi=(41.8\pm 1.2)^\circ$           & 32.8/6\\[1ex]
$f_q=f_\pi$             & $\phi_s=(41.6\pm 1.3)^\circ$      &   &
$f_q=f_\pi$             & $f_s=(1.68\pm 0.07) f_\pi$          &   \\[1ex]
                                 & $f_s=(1.69\pm 0.07) f_\pi$          &       &
                                 &                   &   \\[1ex]\hline
&&&&&\\
$\phi_q$ and $\phi_s$ free  & $\phi_q=(41.6\pm 2.3)^\circ$      & 17.9/4&
$\phi_q=\phi_s\equiv\phi$   & $\phi=(41.5\pm 1.2)^\circ$           & 17.9/5\\[1ex]
$f_q$ free              & $\phi_s=(41.5\pm 1.4)^\circ$      &       &
$f_q$ free              & $f_q=(1.09\pm 0.03) f_\pi$         &   \\[1ex]
                                & $f_q=(1.09\pm 0.03) f_\pi$         &       &
                                & $f_s=(1.68\pm 0.07) f_\pi$         &   \\[1ex]
                                & $f_s=(1.68\pm 0.07) f_\pi$        &       &
                                &                   &   \\[1ex]
\hline\hline
\end{tabular}
}
\caption{Results for the $\eta$-$\eta^\prime$ mixing parameters 
in the octet-singlet basis \textit{(upper part)} and the quark-flavour basis \textit{(down part)}
of the two mixing angle scheme \textit{(left)} and the one mixing angle scheme \textit{(right)}.}
\label{table1}
\end{table}
As seen from Table \ref{table1},
a significant improvement in the $\chi^2/\textrm{d.o.f.}$~is achieved when the constrain
$\theta_8=\theta_0\equiv\theta$ is relaxed
(in the most favorable case the $\chi^2/\textrm{d.o.f.}$~is reduced by a factor of 3),
allowing us to show explicitly in the octet-singlet basis
the improvement of our analysis using the two mixing angle scheme
with respect to the one using the one mixing angle scheme.
Concerning the two mixing angle scheme,
the $\theta_8$ and $\theta_0$ mixing angle values are different at the $3\sigma$ level,
in accordance with the expectations of large $N_c$ Chiral Perturbation Theory (ChPT)
where the difference between the two mixing angles is due to a $SU(3)_f$-breaking effect
proportional to $f_K^2-f_\pi^2$ \cite{Leutwyler:1997yr}.
When $f_8$ is left free the experimental data seem to prefer a value higher than the one
predicted by ChPT ($f_8=1.28 f_\pi$),
while for the parameters $\theta_8, \theta_0$ and $f_0$ our values are in agreement with
those of Ref.~\cite{Leutwyler:1997yr}.
In the one mixing angle case to leave $f_8$ free does not make any substantial difference.

The same kind of analysis can be performed in the quark-flavour basis.
The results are presented in the down part of Table \ref{table1}.
As seen, there is no significant difference at the present experimental accuracy between the
$\chi^2/\textrm{d.o.f.}$ of the fits when data are described in terms of 
two mixing angles (in the quark-flavour basis) or if $\phi_q=\phi_s\equiv\phi$ is imposed.
This behaviour can again be understood in the framework of large $N_c$ ChPT since the difference
of the mixing angles in this basis is proportional to an OZI rule violating parameter which appears
to be small \cite{Feldmann:1999uf}.
It is however important to notice that the fit considerably improves when the parameter $f_q$ is left free.
In this case the value obtained is incompatible with the value in the large $N_c$ limit $f_q=f_\pi$.
The value for the mixing angle $\phi$ agrees with previous phenomenological analyses
\cite{Feldmann:1999uf}, while the large value of $f_s$ (along with a large $f_8$)
is because of the inclusion in our fits of the most recent $\phi\to\eta^\prime\gamma$ decay width.
In all cases, to leave the vector mixing angle free does not produce any effect on the fits.

At the present accuracy, our results satisfy the approximate relations existing between
the two different sets of mixing parameters (only valid for $\phi_q=\phi_s\equiv\phi$)
\cite{Feldmann:1999uf}:
\begin{equation}
\label{mixparosqf}
\begin{array}{ll}
f_8=\sqrt{1/3 f_q^2+2/3 f_s^2}\ ,\quad & \theta_8=\phi-\arctan(\sqrt{2}f_s/f_q)\ ,\\[2ex]
f_0=\sqrt{2/3 f_q^2+1/3 f_s^2}\ ,\quad & \theta_0=\phi-\arctan(\sqrt{2}f_q/f_s)\ .
\end{array}
\end{equation}

\section{Conclusions}
We have performed an updated phenomenological analysis on various decay processes
using the two mixing angle description of the $\eta$-$\eta^\prime$ system with two aims:
\emph{i)} 
to check the validity of the two mixing angle scheme and its improvement
over the standard one angle picture and
\emph{ii)}
to test the sensitivity of the analysis to the mixing angle schemes:
octet-singlet basis vs.~quark-flavour basis.
The agreement between our theoretical predictions and the experimental values
is remarkable and can be considered as a consistency check of the whole approach.
We have shown that a two mixing angle description in the octet-singlet basis is required in order to achieve good agreement with experimental data.
On the contrary, in the quark-flavour basis and with the present experimental accuracy a
one mixing angle description of the processes is still enough to reach agreement.


\begin{thebibliography}{99}
\bibitem{Leutwyler:1997yr}
H.~Leutwyler,
Nucl.\ Phys.\ Proc.\ Suppl.\  {\bf 64} (1998) 223
[arXiv:hep-ph/9709408].

\bibitem{Escribano:2005qq}
 R.~Escribano and J.~M.~Frere,
 JHEP {\bf 0506} (2005) 029
 [arXiv:hep-ph/0501072].
  
\bibitem{Ball:1995zv}
P.~Ball, J.~M.~Frere and M.~Tytgat,
Phys.\ Lett.\ B {\bf 365}, 367 (1996)
[arXiv:hep-ph/9508359].

\bibitem{Eidelman:2004wy}
S.~Eidelman {\it et al.}  [Particle Data Group Collaboration],
Phys.\ Lett.\ B {\bf 592} (2004) 1.

\bibitem{Feldmann:1999uf}
T.~Feldmann,
Int.\ J.\ Mod.\ Phys.\ A {\bf 15} (2000) 159
[arXiv:hep-ph/9907491].
\end{thebibliography}
\end{document}